\documentclass {article}
\usepackage{graphicx}
\usepackage{natbib}

\bibpunct{(}{)}{;}{a}{,}{;}

\newcommand{\be}{\begin{equation}}
\newcommand{\ee}{\end{equation}}

\begin{document}

\title{Shear thickening in densely packed suspensions of spheres and rods confined to few layers}
\author{Eric Brown$^1$, Hanjun Zhang$^2$, Nicole A. Forman$^{2,3}$, Benjamin W. Maynor $^3$,\\
Douglas E. Betts$^2$, Joseph M. DeSimone$^{2,3}$, Heinrich M. Jaeger$^1$ \\
\\
$^1$James Franck Institute, The University of Chicago, Chicago, IL 60637\\
$^2$Department of Chemistry, University of North Carolina, Chapel Hill, NC 27599\\
$^3$Liquidia Technologies, Research Triangle Park, NC 27709}
  

 \maketitle
 
\begin{abstract}

We investigate confined shear thickening suspensions for which the sample thickness is comparable to the particle dimensions.  Rheometry measurements are presented for densely packed suspensions of spheres and rods with aspect ratios 6 and 9. By varying the suspension thickness in the direction of the shear gradient at constant shear rate, we find pronounced oscillations in the stress.  These oscillations become stronger as the gap size is decreased, and the stress is minimized when the sample thickness becomes commensurate with an integer number of particle layers.  Despite this confinement-induced effect, viscosity curves show shear thickening that retains bulk behavior down to samples as thin as two particle diameters for spheres, below which the suspension is jammed.  Rods exhibit similar behavior commensurate with the particle width, but they show additional effects when the thickness is reduced below about a particle length as they are forced to align; the stress increases for decreasing gap size at fixed shear rate while the shear thickening regime gradually transitions to a Newtonian scaling regime.    This weakening of shear thickening as an ordered configuration is approached contrasts with the strengthening of shear thickening when the packing fraction is increased in the disordered bulk limit, despite the fact that both types of confinement eventually lead to jamming.   

\end{abstract}

When fluids are confined to thin layers, their flow behavior can differ markedly from the bulk.  This is an issue of considerable importance in situations ranging from molecular lubrication films  where it can induce increased friction \citep{BN06} to macroscopic granular materials flowing out of a narrow hopper opening, where the particles can jam into rigid structures.  Here we investigate this transition from flowing to jamming for densely packed, non-Brownian suspensions which exhibit shear thickening in the absence of strong interparticle interactions \citep{Ba89, BFOZMBDJ10}.  Shear thickening fluids are non-Newtonian such that their dynamic viscosity -- defined as shear stress divided by shear rate in a steady state -- increases over some range of shear rate.  In dense suspensions this phenomenon is remarkable because it is characterized by a dramatic increase of stress with shear rate \citep{MW58, Ho72,Ho82, La94,FHBM96,MW01a, EW05,LDHH05, FHBOB08, BJ09} which goes by the name of Discontinuous Shear Thickening, as well as the ability to absorb impacts \citep{LWW03}.   Consequences of jamming can be seen even in bulk rheology in terms of a critical packing fraction $\phi_c$ corresponding to random loose packing above which the suspension is jammed, i.e. a yield stress is measured.  This critical packing fraction controls the shear stress as a function of shear rate such that the slope increases with packing fraction and becomes discontinuous at $\phi_c$ \citep{BJ09}.  Prior work on shear thickening fluids was mostly limited to bulk behavior.  Here we take the opposite approach and study shear thickening suspensions in the limit of only a few layers thick.

A key quantity in fluid-jammed transitions of granular systems is the system size range over which the transition occurs.  Considering that differently shaped particles can arrange into different structures, we investigate this transition for both spheres and long rods.  While the angular orientation of spheres with respect to the velocity gradient is irrelevant to the flow structure, the orientation of long rods is likely to be very significant.  In thin, highly confined samples, rods can be forced to align in the plane of the sample, and this can be used to investigate which particle degrees of freedom are important to shear thickening.  The potential for different  transition behavior is especially intriguing considering that shear thickening in bulk is qualitatively similar for suspensions of spheres and rods \citep{MW01a,EW05}.  

We first address the fluid-jammed transition under confinement in Sec.~\ref{sec:gapsweep}.  We performed fixed-shear rate rheological measurements for  suspensions of spheres and rods that are only a few layers thick in the direction of the shear gradient.  We show that there is a confinement effect such that the stress varies non-monotically with the gap size and has local minima commensurate with integer numbers of particle layers.  This confinement effect is measurable up to about nine layers of spheres, and down to two layers, below which the system is jammed. Additionally, the rods  are forced to align in small gaps and show an increase in stress at gap sizes below about a particle length as particle rotation in the plane made by the shear direction and gradient is suppressed.  We then show that the onset of jamming occurs at 2 particle layers over a wide range of packing fractions.  In Sec.~\ref{sec:shearthickengap} we show full viscosity curves to determine how shear thickening behavior evolves at small gaps.  We first address this for spherical particles.   While we find transient jamming in stress-controlled measurements for small gaps, remarkably, bulk shear thickening behavior remains unchanged down to two particle layers.  Below two particle layers the system becomes jammed.  For long rod-shaped particles we find the same jamming below two particle layers, but there is an additional gradual transition.  As the gap size is reduced below about a particle length the shear thickening weakens, approaching a Newtonian scaling regime before the system jams.  In the final section, we contrast these results with the transition from shear thickening to jamming at $\phi_c$ in bulk systems and discuss them in terms of different proposed mechanisms for shear thickening.

\section{Materials and methods}
\label{sec:methods}

We used two types of particles to explore geometric effects: spheres and rods.  To ensure geometric effects were prominent we worked with particles above 10 $\mu$m in size to minimize effects of Brownian motion and interparticle interactions.  The spheres were smooth soda-lime glass (density 2.46 g/mL) obtained from MoSci corporation (Class IV) with a hydrophobic silane coating.  The nominal size range was such that 80\% of particles pass through a 120 mesh sieve and are retained by a 170 mesh sieve.  The particle diameters were manually measured with an optical microscope; we found a mean diameter $a=89\pm 2$ $\mu$m (half of this uncertainty is absolute from our measurement scale and half is statistical based on many particle measurements) with a standard deviation of 12 $\mu$m.  The peak of the size distribution is $92\pm3$ $\mu$m and skewed slightly to smaller sizes, part of which is due to broken pieces which make up about 3\% of the particles.  These spheres were dispersed in mineral oil with a viscosity of 58 mPa$\cdot$s and density of 0.87 g/mL.  The hydrophobic coating was used to minimize the particle-fluid surface tension.  The value of this surface tension must be below a threshold for a suspension to exhibit shear thickening \citep{Ba89, BFOZMBDJ10}.  However, we note that uncoated glass spheres will also shear thicken in mineral oil, so this surface tension was already small enough and the coating was not necessary to obtain shear thickening.   

\begin{figure}                                                
\centerline{\includegraphics[width=5.75in]{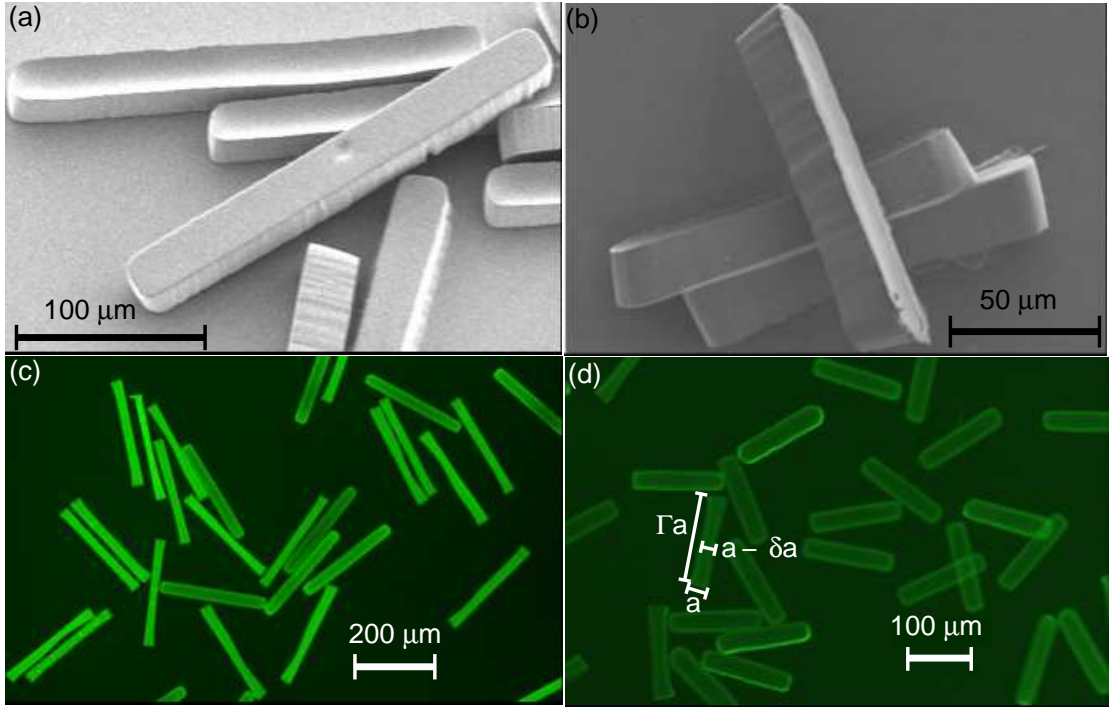}}
\caption{Images of PEG rods (a,b): dry particles viewed with an scanning electron microscope.  (c,d):  particles soaked in PEG viewed with an optical microscope with fluorescence. (a,c): aspect ratio $\Gamma=9$.  (b,d): $\Gamma=6$.   Dimensions of particles are illustrated in panel (d). }                                        
\label{fig:rodimages}
\end{figure}

The rods were fabricated using the PRINT$^{\textregistered}$ process \citep{RMEEDD05} to control the shape of the particles.   We designed the chemistry specifically to minimize particle-fluid surface tension so the suspension would exhibit shear thickening.  Typically, the monomer solution was prepared as follows: 0.01 g of 1-hydroxycyclohexyl phenyl ketone (HCPK, Aldrich), and 0.02 g of fluorescein o-acrylate (Aldrich) were placed into an Eppendorf tube followed by the addition of 0.1 ml of N,N-dimethylformamide (DMF, Aldrich). The monomer mixture was then mixed thoroughly by vortex mixing to dissolve the HCPK photoinitiator and the fluorescein o-acrylate fluorophore.  Lastly, 0.57 g of trimethylolpropane ethoxylate triacrylate (Mn = 912 g/mol, Aldrich) and 0.40 g of poly(ethylene glycol) methyl ether acrylate (Mn = 454 g/mol, Aldrich) were added to the monomer mixture and vortex mixed again. The resulting solution was composed of 57\% (w/w) triacrylate, 40\% (w/w) methyl ether acrylate, 1\% (w/w) HCPK, and 2\% (w/w) fluorescein o-acrylate.  Rectangular rods were then fabricated from this solution using the PRINT process in two sets with different  aspect ratios $\Gamma$.  The rods were suspended in poly(ethylene glycol) dimethyl ether (PEG) (Mn = 250g/mol, Aldrich).   The dimensions of particles that had been soaked in PEG were measured with an optical microscope.  One set had mean dimensions of ($266.5 \pm0.6$ $\mu$m)$\times (29.9\pm0.5$ $\mu$m)$\times (31.4\pm0.4$ $\mu$m).  The standard deviation in each particle dimension from particle to particle is 2-3 $\mu$m.   The shortest length is non-uniform for individual particles as can be seen in Fig.~\ref{fig:rodimages}.  The width of $a=29.9$ $\mu$m listed above is for the mean width at the ends of the particle.  On average the width is less in the center of the particles by $\delta a =10.6$ $\mu$m than at the ends.  The aspect ratio of these particles, defined as the longest dimension divided by the shortest dimension $a$, is $\Gamma = 8.9$ (we will refer to these as $\Gamma=9$ rods).  The other set had mean dimensions ($139.5\pm0.7$ $\mu$m)$\times(24.9\pm0.6$ $\mu$m)$\times(30.2\pm0.4$ $\mu$m).  In this case $a=24.9$ $\mu$m and $\delta a = 8.2$ $\mu$m.  These particles have aspect ratio $\Gamma=5.6$ (we will refer to these as $\Gamma=6$ rods).  Because of the large aspect ratio of these particles they have a tendency to break under stress and about 18\% of the $\Gamma=9$ rods and 9\% of the $\Gamma=6$ rods were found broken after being sheared in rheology experiments.

Measurements were performed with an Anton Paar Physica MCR 301 rheometer which measures the torque $T$ required to shear a sample at a tool rotation rate $\omega$.  A parallel plate setup was chosen so that the sample thickness $d$ in the horizontal gap between plates could be varied continuously.  The plate radius used was $R=12.5$ (or $25$) mm.  The effective viscosity, indicating the mechanical resistance to shear, is defined as $\eta\equiv\tau/\dot\gamma$ in a steady state for shear stress 

\be
\tau = \frac{2T}{\pi R^3}
\label{eqn:stress}
\ee

\noindent and shear rate 

\be
\dot\gamma = \frac{R\omega}{d} \ .
\label{eqn:shearrate}
\ee

\noindent  These definitions are meant to characterize the mechanical response; we do not imply a linear shear profile.  The thickness can be set with a resolution of 1 $\mu$m and with a parallelism of $\pm3$ $\mu$m.   The plate surfaces are smooth and the tool surfaces are stainless steel.  

Measurements were made at a bottom plate temperature controlled at $20^{\circ}$ C with the room humidity ranging from 22\% to 38\%, although during individual experiments the humidity was constant.  This affects solvent evaporation/adsorption which can have a significant effect on the rheology due to the sensitive packing fraction dependence of shear thickening suspensions \citep{BJ09}.  Care was taken so that no fluid extended outside the parallel plates and the particles were confined to the space between the plates by surface tension.   This boundary condition means that dilation under shear can result in normal stresses on the rheometer tool and has a significant effect on discontinuous stress-shear rate curves \citep{FHBOB08}.  Samples were pre-sheared immediately before measurements for at least 100 seconds at shear rates above the shear thickening regime where the steady state flow is fully mobilized.  After this pre-shear, measurements on suspensions were found to be repeatable with a typical variation of 10-20\% from run to run.   This is the variation whether we remeasure a sample that is in place on the rheometer or replace it with a new one with the same procedure.   Following the preshear, measurements were performed with decreasing and then increasing control parameter (stress, shear rate, or gap size, depending on the test) and additional runs were made with different control ramp rates to check for hysteresis, thixotropy, and transients.  Here we usually show only one set of curves for brevity if they were all identical within typical variations.  Since glass spheres are denser than mineral oil, we checked that the bulk shear thickening behavior is qualitatively the same if the spheres are instead dispersed in a density matched fluid at $d=500$ $\mu$m.  The rods are nearly density matched with the PEG solvent they are dispersed in.  The $\Gamma=9$ rods at an initial packing fraction $\phi = 0.26$ were measured to settle at a terminal rate of 4 $\mu$m/s before geometric constraints arrest them.  This settling rate is below the plate edge speed in the shear thickening regime for all measurements with the rods. For reference, this settling rate would be comparable to that of 89 $\mu$m glass spheres if they had a 1\% density mismatch with their suspending fluid.  We have measured bulk shear thickening in the sphere suspensions with both rough and smooth plates and did not find any difference in the shear thickening due to the plate surface.   We avoided rough plates in the following experiments because they could be responsible for additional jamming effects at small gaps.  To directly measure slip, we used video microscopy to observe the shear profile at the outer edge of the plate.  The results of these measurements are shown in the Appendix and confirm slip is not responsible for the measured gap size dependence.

\section{Gap size sweep at fixed shear rate}
\label{sec:gapsweep}

We first investigate the transition from fluid to jammed states for thin layers of suspensions at constant shear rate.  To do this we measured the shear stress and normal stress on the top plate as the rheometer gap size was reduced down to a few particle layers.  We performed this sweep at constant shear rate to study the gap size dependence only and we will consider the effect on shear thickening later.  To vary the gap continuously, the initial radius of the sample was made smaller than that of the plates so when the gap size was reduced the sample could expand horizontally without spilling.   To obtain the stress and shear rate of the sample we use Eqns.~\ref{eqn:stress} and \ref{eqn:shearrate} but replace the plate radius $R$ with the radius of the sample which varies as $R = \sqrt{V/(\pi d)}$ for a constant volume $V$.  We measured the stress as the gap size was ramped down and then back up with a rotation rate varying as $\omega \propto d^{1.5}$ so that the shear rate $\dot\gamma = \omega R/d$ remained constant. 

\begin{figure}                                                
\centerline{\includegraphics[width=6.in]{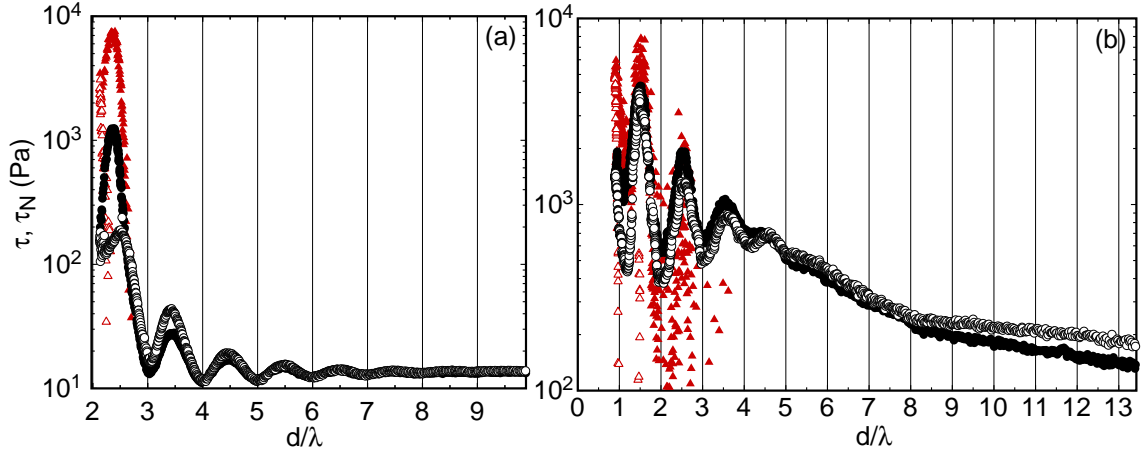}}
\caption{Shear and normal stresses as a function of gap size $d$ at fixed shear rate.  The gap size is scaled by the measured wavelength $\lambda$ of oscillations.  (a) spheres of diameter 89 $\mu$m and $\lambda = 86.1$ $\mu$m.  (b) aspect ratio $\Gamma=9$ rods of width $a=29.9$ $\mu$m and $\lambda=29.8$ $\mu$m.  Solid circles:  shear stress $\tau$ for decreasing gap.  Open circles:  $\tau$ for increasing gap.  Solid triangles (red online): normal stress $\tau_N$ for decreasing gap.  Open triangles (red online):  $\tau_N$ for increasing gap.}
\label{fig:gapsweep}                                        
\end{figure}

We show data first for a suspension of glass spheres in mineral oil at $\phi=0.50$ (below $\phi_c$).  The sample started out with a radius of 10 mm at a gap height of $d= 1000$ $\mu$m for a 25 mm radius plate with a shear rate $\dot\gamma= 4$ s$^{-1}$.  The gap dependence of the stress was measured several times consecutively to ensure the rheometer plates were wetted and to check for repeatability.  Stress measurements are shown in Fig.~\ref{fig:gapsweep}a.     Tests were also run at different shear rates and with different rates of gap size change to check that the observed effects occurred at consistent gap sizes and did not depend on tool orientation or temporal oscillations.   We find that the shear stress approaches a constant value at large gaps as expected for bulk fluid behavior.  As the gap decreases an oscillation of increasing amplitude is superimposed on that background.      Based on the oscillation amplitude this confinement-induced effect is quite strong for layers 2-3 particles thick but becomes about 10\% of the measured stress at 6-7 particle layers, at which point this gap size dependence would not be resolvable in viscosity curves from run to run.  Analysis of the wavelength and phase of these oscillations is shown in Sec.~\ref{sec:gapsweepanalysis} to shed light on the structure of these suspensions at small gaps.  The absolute normal stress $\tau_N$ is plotted in Fig.~\ref{fig:gapsweep}.  Typically positive normal stresses are generated along with discontinuous shear thickening \citep{LDHH05, FHBOB08}, but the net normal stress in these measurements is negative at larger gap sizes where shear-induced stress is relatively low and the dominant effect comes from surface tension which is observed even at zero shear rate.

We performed the same experiment with a suspension of $\Gamma=9$ rods at $\phi =0.30$ (below $\phi_c$) and $\dot\gamma =31$ s$^{-1}$.  At an initial gap height of 720 $\mu$m the sample diameter was 4.5 mm for a 50 mm diameter plate.  While both spheres and rods showed commensurability effects at small gap sizes indicated by the oscillations in Fig.~\ref{fig:gapsweep}, the major difference between the particle shapes  is the baseline behavior.  There is a kink in Fig.~\ref{fig:gapsweep}b below which the stress increases more dramatically as the gap size is reduced.  This kink was reproducible when experiments were repeated several times.   This change in scaling occurs when the gap is roughly equal to one particle length (in units of $d/a$ as in Fig.~\ref{fig:gapsweep} this length is equal to the aspect ratio) and does not occur for spherical particles.  This suggests that the additional particle length scale of rods can be observed in rheological measurements.  In this case the stress becomes higher as the small gap prevents particle rotation out of the horizontal plane.  

 We note some other observations based on performing multiple runs.  The measurements shown in Fig.~\ref{fig:gapsweep} were done just below $\phi_c$ and at shear rates just above the shear thickening regime in each case as can be confirmed by comparing the stress in the bulk limit to Figs.~\ref{fig:gapsize_spheres} and \ref{fig:Gamma12}.  We chose different values that are in the same rheological regime rather than at the same shear rate and packing fraction which would have been in different rheological regimes due to the different shapes.  We confirmed that the shear flow is fully developed (i.e.~all particles are in motion with a nearly linear shear profile) in this regime using video microscopy (see Appendix).  We also performed measurements at lower shear rates in a region where the shear flow is not fully developed; in those cases oscillations in the stress were less apparent if observed at all.  This suggests that these confinement-induced effects are less prominent at stresses below the shear thickening regime where the shear flow is not fully developed.  Similar oscillations were seen above $\phi_c$ where there is a large yield stress, but the background increased as the gap is decreased even for spheres instead of remaining flat for large gaps.  A reproducible kink was never observed in the background for spheres.  The peak heights are somewhat variable from run to run, and there is no clear preference for peaks from either decreasing or increasing gap measurements to be larger except for the first peak (i.e. the 2nd peak being larger for increasing gap in Fig.~\ref{fig:gapsweep}a is not a reproducible feature).  For spheres, the first peak seems to be significantly affected by hysteresis at the turnaround point which most apparent in the normal stress measured on the top rheometer plate.  The normal stress shown in Fig.~\ref{fig:gapsweep}a became measurable during the lowest peak and approached $24$ kPa (the maximum our rheometer could handle) as the gap size was reduced down to 183 $\mu$m ($2.06a$).  For this reason we could not compress the sample to smaller gap sizes.  When the gap size was then increased from this point, the normal stress dropped off much more quickly than it grew for decreasing gap size.  This hysteresis effect suggests that as the gap decreases and forces the particles to arrange into fewer layers, this rearrangement is met with much resistance.  As the gap size was increased, there is much less resistance.  Through friction this normal stress can translate to a shear stress, resulting in much more shear resistance for decreasing gap size than for increasing gap size when the normal force is present, as seen in the first peak of Fig.~\ref{fig:gapsweep}a where the normal stress exceeds the shear stress.  This hysteresis effect is also likely the reason the first pair of peaks is slightly offset with the peak for decreasing gap to the left and the peak for increasing gap to the right.

\subsection{Structures under confinement} 
\label{sec:gapsweepanalysis}

\begin{figure}                                                
\centerline{\includegraphics[width=5.in]{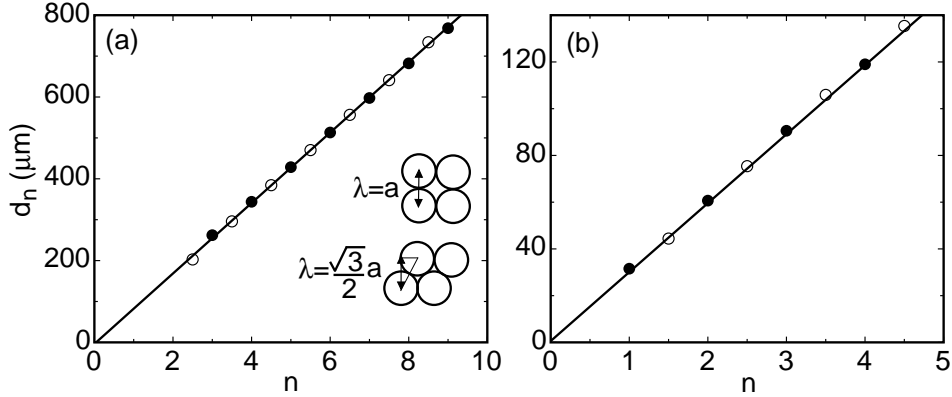}}
\caption{The extrema locations $d_n$ vs. extrema number $n$ from oscillations of the stress as a function of gap size.  (a) spheres with diameter $a=89$ $\mu$m.  Insets: diagrams of possible layer arrangements for shear flow into the page and their associated wavelengths.  The measured wavelength is consistent only with non-overlapping layers as in the upper diagram, although organization within those layers is not implied.  (b) $\Gamma=9$ rods with width $a= 29.9$ $\mu$m.    Solid circles:  local minima of the stress as a function of gap size.  Open circles: local maxima. }
\label{fig:gapsweepphase}                                        
\end{figure}

The wavelength and phase of the confinement-induced oscillations can give some insight into the structure of the shear flows in thin layers.  The oscillations do not fit well to standard decaying sinusoidal functions, so to identify the extrema we fit Gaussian functions locally to each extrema to obtain the gap size $d_n$ at each valley (which we assign to integer $n$ in order so that smaller $n$ corresponds to lower gap size) and peak (which we assign to integer plus $1/2$ values of $n$).  These are then plotted in Fig.~\ref{fig:gapsweepphase} vs. peak number $n$ for decreasing stress measurements only.  The extrema appear to be equally spaced as expected for oscillations with a fixed wavelength.  A linear fit $d_n = (\Psi+n)\lambda$ is used to obtain the wavelength $\lambda$ and phase $\Psi$ (in revolutions) of the oscillations.\footnote{There is an arbitrary integer offset in $n$ which we choose so that extrapolations of the extrema locations for $d_n=0$ appear at $|n| < 1/2$ for ease of interpretation of the phase offset (since the oscillations are periodic phase offsets by integer revolutions are equivalent).}  

For the spheres we obtain $\lambda = 86.1\pm 0.5$ $\mu$m and $\Psi = -0.04 \pm 0.04$ (with statistical uncertainties from the fit).  The wavelength is compatible with the mean particle size of $89\pm2$ $\mu$m, and well within a standard deviation of 11 $\mu$m of the mean particle size.  Because of the relatively large polydispersity, it is apparent that the commensurability of the gap size depends on the mean particle size rather than the maximum particle size, for example.   Since the stress minima occur at integer multiples of the wavelength, this indicates that flow is easier when the particles fit nicely into layers.  At intermediate gaps particles cannot arrange into layers so are more likely to be forced into arrangements where particles cannot easily shear past each other.   

Previous studies have shown that confined sphere packings can under certain conditions order into crystalline structures, for example with confinement in the direction normal to the shear plane \citep{CMW04} or without shear \citep{TGR02, DW09}.  Some crystalline structures would result in different wavelength and phase values.  For example, if the spheres were hexagonally close packed a zig-zag sliding motion could take advantage of the larger gaps above and between particles \citep{VS05} instead of going directly over top of each other.  This sort of structure could result in a wavelength as small as $\sqrt{3}a/2$ and a phase difference of $\Psi = 1 - \sqrt{3}/2$ because the smaller thickness per layer cannot be taken advantage of next to the top and bottom plates.   Any flow structure that is arranged into overlapping layers would result in $\lambda<a$ and $\Psi \ne 0$.   The likely significance of the measured values of $\lambda=a$ and $\Psi=0$ is that such non-overlapping layers can still shear past each other but do not have to have any ordered structure within the layer, so this is the only wavelength and phase compatible with disorganized shear and so is entropically favorable.

For $\Gamma=9$ rods we fit $d_n(n)$ to obtain $\lambda = 29.8\pm0.2$ $\mu$m.  This wavelength is in the range of the two short dimensions of the rods suggesting it is set by one of those dimensions.  Within measurement uncertainties it is consistent with the shortest dimension measured at the ends of the particles, $a= 29.9\pm0.6$ $\mu$m.    While we could have chosen a number of different values to characterize the particle width $a$ given the variation, none of the other obvious choices are consistent with $\lambda$.  For example the intermediate length is $31.4\pm0.4$ $\mu$m,  the shortest width averaged over the length of the particle $a-\delta a/2$ is $24.6\pm0.5$ $\mu$m, and the average of these two is $28.0\pm0.6$ $\mu$m.  The likely significance of the value of $a$ is that it corresponds to the mean value of the smallest channel that particles could slide through.  These oscillations with this length scale suggest a tendency towards alignment of rods with the shear direction and enforced layering of particles in their thinnest dimension as the gap size is reduced.  The fit of $d_n(n)$ gave the phase offset $\Psi = 0.03\pm0.02$, again indicating that the stress is minimum and it is easier to flow for an integer number of layers.  

For both spheres and rods, shear is easiest when the gap size is commensurate with an integer number of particle layers.   The measured wavelengths and phases of oscillations in both cases suggested that the particles arrange in such a way that they can shear in non-overlapping layers.  While the rods exhibited a higher background stress as particle rotation was suppressed at gaps smaller than a particle length, the spheres have no such geometric restrictions on rotation and exhibited a uniform background stress.

\subsection{Onset of jamming in confined systems}

We observed that suspensions of both spheres and rods cannot be further compressed beyond about two particle layers.  In the bulk limit the onset of jamming, defined by the onset of a yield stress, occurs when the packing fraction is increased in a disordered system \citep{OSLN03}.  One approach to understand the significance of jamming at two layers may be to interpret the confinement effect as a deviation of the packing fraction at the onset of jamming from the bulk value.  For example, this packing fraciton is known to decrease in systems of frictionless spheres that are reduced to less than about ten particle layers \citep{DW09}.  

  \begin{figure}                                                
\centerline{\includegraphics[width=5.in]{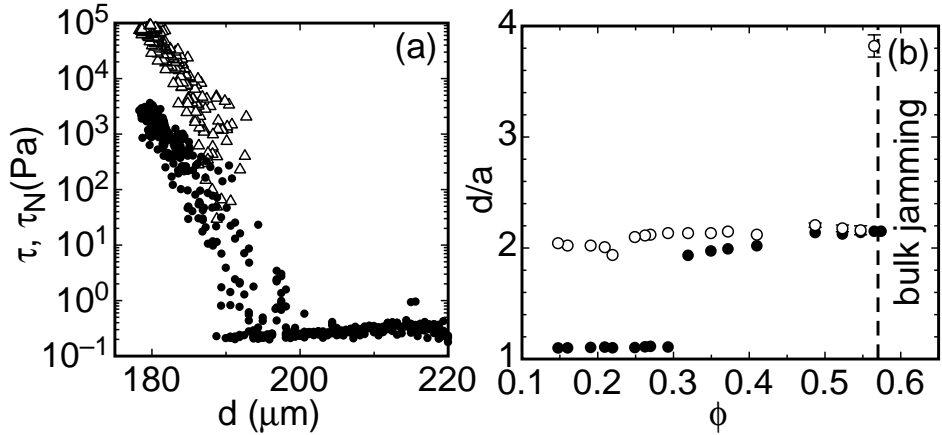}}
\caption{(a)  Shear stress $\tau$ (solid circles) and normal stress $\tau_N$ (open triangles) as a function of gap size for 89 $\mu$m glass spheres at a packing fraction $\phi=0.41$.  (b) Jamming phase diagram showing the transition gap size $d/a$ normalized by mean particle size for different $\phi$.  Open sysmbols: the onset gap size of a yield stress for each packing fraction, below which the system is jammed.  Solid symbols: the minimum gap that could be reached at 40 kPa normal stress. Dashed line:  $\phi_c$, above which the system is jammed in the bulk limit.}  
\label{fig:jamphase_gap}                                        
\end{figure}

To determine the onset of jamming in confined systems, we measure the shear and normal stress at a fixed shear rate of 1 s$^{-1}$, and reduce the gap size at a rate of 0.1 $\mu$m/s.  We do this for a wide range of packing fractions $\phi$.  This is similar to the previous measurements, except now we are working in the limit of small shear rates to measure the onset of jamming.    A sample pair of stress curves is shown in Fig.~\ref{fig:jamphase_gap}a.  The onset of jamming is taken as the smallest gap size where the shear stress remains at a low value before it increases dramatically as the gap size is reduced.  This is plotted in Fig.~\ref{fig:jamphase_gap}b for different packing fractions.  Additionally, we plot the minimum gap size reached, obtained when the compressive stress reached 40 kPa (the maximum applied).  This is similar to the condition for the minimum size reached in the experiments shown in Fig.~\ref{fig:gapsweep}, although those measurements were made at much higher shear rates.  To check that this is indeed a yield stress in the limit of quasistatic compression, we varied the shear rate from $10^{-3}$ to 10 s$^{-1}$ and and compression strain rate from to $3\times10^{-5}$ to $2\times10^{-2}$ s$^{-1}$.  Fluid and jammed states could be distinguished because the low stress value varied with the shear rate such that the viscosity was the same at different shear rates, and the high stress values at small gaps remained independent of shear rate as in the case of a yield stress.   At small $\phi$, there was no significant change in the gap size at the transition or the minimum gap size reached when varying the shear or compression rate.  At $\phi\stackrel{>}{_\sim}0.5$, the minimum gap value increased for increasing compression rate or decreasing shear rate, suggesting that a large shear is important for driving rearrangement of particles under compression so they can pack more efficiently.   In every case the minimum gap values approached an asymptote in the limit of low compression rate and high shear rate, and the plotted values are consistent with that limit \footnote{Even though this is not technically the zero-shear rate limit because it must be large compared to the compression rate, the stress is not increasing with shear rate so this still effectively measures a yield stress.}.  Near the jamming transition at small gaps, there are many large fluctuations the stresses.  These values are not a reproducible function of the gap size upon repeated measurements, indicating they are transient fluctuations.

It is seen that the onset of jamming remains constant at about $2.1a$ over a wide range of packing fractions.  It is likely that this is somewhat larger than $2a$ because the polydispersity  in particle size is about 13\%, so in some places the layers will be thicker than others.  Very near the bulk jamming transition at $\phi_c$, the onset of a yield stress occurs at a higher gap size.  This is a sharper transition than seen in simulations of frictionless spheres with periodic boundaries in two directions \citep{DW09}.  The difference in packing fraction at the onset of jamming is usually attributed to friction \citep{JSSSSA08}.

We note that there is a hysteresis effect in the onset of the yield stress.  For example at $\phi=0.2$, if we compress to $d=1.3a$, then lift the plate up to $d=1.45a$ the normal force drops immediately, and upon shearing there is no yield stress.  In other words, the onset of a yield stress is lower for a history of increasing gap size compared to a history of decreasing gap size.  We also note that the yield stress grows slowly over a wide range of gap size at these small packing fractions.  If the yield stress was coming from the compression of hard spheres with a Hertzian contact law \citep{OSLN03}, the compressive strain of individual particles at that stress would be on the order of $10^{-3}$, much too little to account for a compression of about 50\% past the onset of the yield stress.  Thus the yield stress must be coming from a different source, perhaps friction, or capillary forces from the generation of trapped air bubbles created as the packing rearranges catastrophically.  While the maximum applied stress of 40 kPa is an arbitrary value, as can be seen in Fig.~\ref{fig:jamphase_gap}a the minimum gap size reached is not very sensitive to this value.  It is probably a good approximation of the onset of jamming due to the compression of hard particles since we are not aware of other mechanisms that could support such large stresses. 

The minimum gap size reached remains about $2a$ throughout the entire shear thickening range.  While the minimum gap size drops to about $1.1a$ for $\phi <0.30$, we note that Discontinuous Shear Thickening does not occur at such low packing fractions of spheres.  We can understand the sharp transition in minimum gap size based on a free volume argument.  The fact that the minimum gap sizes correspond to approximately integer numbers of layers suggest that the particles are effectively packing into layers at these small gaps and do not pack as efficiently at a small, non-integer numbers of layers.  As the system is compressed below about $2a$ it must transition from 2 layers to effectively 1 layer.  When crossing this transition, the effective free volume that particles can make use of is approximately halved, as confirmed by the fact that the transition from 2 to 1 layers is at about half the packing fraction of the bulk jamming value.  Thus, the significance of jamming at two layers seems to be that there is a large loss of effective free volume. 


\section{Shear thickening at different gap sizes}
\label{sec:shearthickengap}

\subsection{Spheres}

  \begin{figure}                                                
\centerline{\includegraphics[width=3.in]{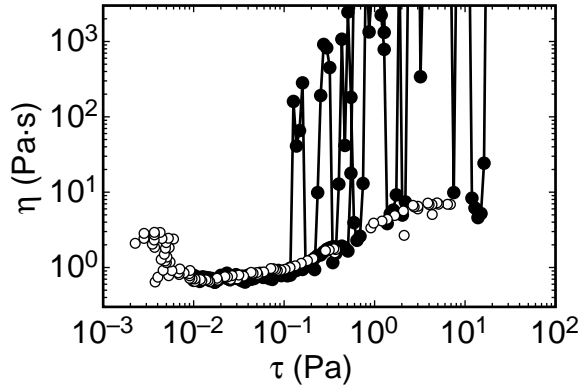}}
\caption{Apparent viscosity curves in controlled shear stress mode (solid symbols) and controlled shear rate mode (open symbols) for spheres at $\phi=0.558$ with a gap size $d=260$ $\mu$m ($2.92a$).  Each vertical line off-scale corresponds to a temporary jam of the rheometer. }  
\label{fig:smallgap_control}                                        
\end{figure}

We next measured viscosity curves for spheres at small gaps to observe the effect of confinement on shear thickening.  We started with a suspension of spheres at $\phi=0.558$ (below $\phi_c$) at a large gap size and removed some sample each time to get to a lower gap, bringing the plate down so the sample just fills the gap to avoid slop.  At each gap height we measured viscosity curves for 10 s per measuring point.   We performed tests in both stress and shear rate-controlled mode and found agreement for large gaps but a difference appeared for small gaps.  For gaps below about $d=300$ $\mu$m ($=3.4a$) in stress-controlled mode, large apparent variations in viscosity were measured.  A comparison of measurements in stress and shear rate-controlled mode can be seen in Fig.~\ref{fig:smallgap_control} for $d=2.6$ mm ($2.92a$).  When the experiment was repeated, these jumps in stress-controlled mode do not occur at the same stress so they are indicative of large fluctuations.  Excluding these jumps, the underlying viscosity curves agree from run to run.  These apparent high viscosities are the result of dramatic drops in shear rate, many of which are consistent with zero shear rate over the 10 s measuring point duration within the resolution of the measurement of $10^{-4}$ s$^{-1}$.  We interpret these events as temporary jams of the rheometer due to chains of particles locking up across the gap.  If we measure the same range in shear rate-controlled mode these temporary jams are not found.  We obtain typical shear thickening curves that are similar to those at larger gap sizes and match the underlying viscosity curves in the stress-controlled measurement without fluctuations.  In shear rate-controlled mode, the rheometer can simply break those chains of particles apart without getting stuck.  For this reason we could only measure in shear rate-controlled mode at the smaller gap sizes.

  \begin{figure}                                                
\centerline{\includegraphics[width=2.5in]{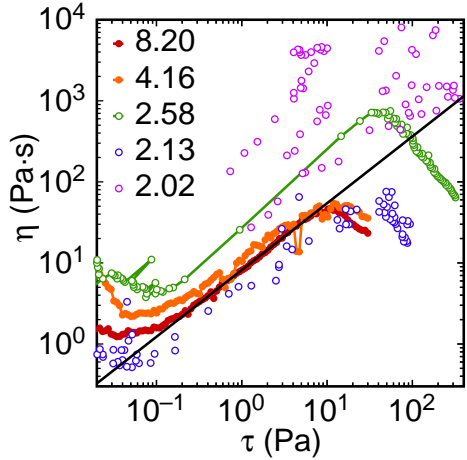}}
\caption{Apparent viscosity curves for different gap sizes for spheres. Solid symbols:  shear stress-controlled data.  Open symbols:  shear rate-controlled data.  The large variations are indicative of fluctuations and not mean behavior.  Legend: values of  gap size $d/a$ in units of particle diameters.}  
\label{fig:gapsize_spheres}                                        
\end{figure}

\begin{figure}                                                
\centerline{\includegraphics[width=3.25in]{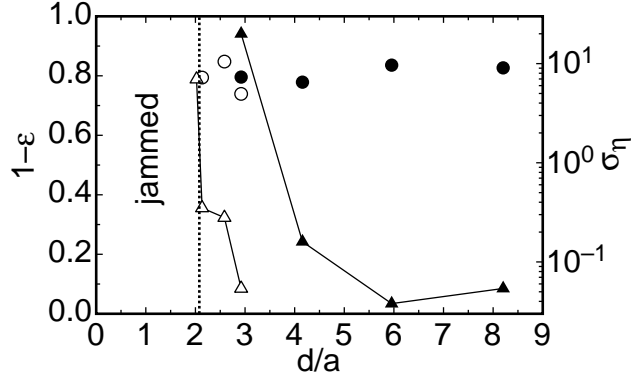}}
\caption{Circles (left axis): Logarithmic slope $1-\epsilon$ corresponding to fits of $\eta\propto \tau^{1-\epsilon}$ to viscosity curves for diffferent sample thicknesses $d$ for $a=89$ $\mu$m diameter spheres.  Triangles (right axis):  fluctuations in viscosity $\sigma_{\eta}$ measured as a fraction of the baseline viscosity.  Solid symbols: shear stress-controlled mode.  Open symbols: shear rate-controlled mode. Dotted line: separates the shear thickening and jammed states. }  
\label{fig:gapsize_epsilon_spheres}                                        
\end{figure}

Viscosity curves can be seen in Fig.~\ref{fig:gapsize_spheres} at different gap sizes for decreasing stress or shear rate ramps.  Even in shear rate-controlled mode measurements still exhibited large fluctuations at the smallest gaps.  While significant fluctuations are seen at a gap of $d=190$ $\mu$m ($2.13a$), a dramatic increase in fluctuations is seen at $d=180$ $\mu$m ($2.02a$), and there is no clear underlying shear thickening curve as is the case at larger gaps.  In shear rate-controlled mode these stress fluctuations were significant enough at $d=180$ $\mu$m that the rheometer motor could not adjust quick enough to keep the shear rate at the set value.  These fluctuations can be quantified by a root-mean-square logarithmic variation $\ln\sigma_{\eta} = \sqrt{\langle (\ln \eta-\ln\eta_{bl})^2\rangle}$.  Here $\eta_{bl}$ is the baseline viscosity curve which is obtained from a fit excluding the large fluctuations since they tend to be rare and positive only.   The value of $\sigma_{\eta}$ indicates a typical fractional variation from the baseline viscosity such that $\sigma_{\eta}=1$ corresponds to a typical variation of 100\% of the baseline value.  These fractional fluctuations $\sigma_{\eta}$ are plotted using the right axis of Fig.~\ref{fig:gapsize_epsilon_spheres} as a function of gap size for both stress- and shear rate-controlled data.  If $\sigma_{\eta}$ becomes comparable to or exceeds unity, then obtaining a meaningful steady state viscosity curve becomes difficult. 

To quantify the effect of gap size on shear thickening we next focus on steady state viscosity curves.  There is not yet a well-accepted quantity for the strength of shear thickening.  Bulk Discontinuous Shear Thickening typically occurs in a relatively fixed stress range and has a slope that increases with packing fraction up to $\phi_c$ \citep{MW01a, EW05, BJ09}.  Shear thickening suspensions seem to have a similar scaling with packing fraction such that $\eta \propto \tau$ in the limit approaching $\phi_c$ from below \citep{BJ09}, so the maximum dynamic range of the viscosity is the same as the dynamic range of the stress.  Thus the stress range of the shear thickening region is a good measure of shear thickening strength for comparing different suspensions.  However, we do not find a reproducible trend in the stress range with gap size for any particle shape.  Alternatively, the slope of the shear thickening region can be used to quantify the strength of shear thickening.  We characterize the shear thickening part of the curve by $\eta \propto \tau^{1-\epsilon}$.  This definition is equivalent to $\tau\propto \dot\gamma^{1/\epsilon}$ and Newtonian flow corresponds to $\epsilon=1$ and a discontinuous jump in the stress-shear rate curve corresponds to $\epsilon=0$.  We fit $\eta \propto \tau^{1-\epsilon}$ to viscosity curves in a fixed stress range of 0.3-1 Pa which corresponds to the steepest portion of the shear thickening region.    Values of the slope $1-\epsilon$ vs. gap size are shown in Fig.~\ref{fig:gapsize_epsilon_spheres}.  The variation from run to run can be attributed to the high sensitivity of the magnitude and slope of the viscosity curves to the packing fraction which can change from run to run as sample is removed.  At this packing fraction,  a 1\% change in packing fraction can cause a change in $\epsilon$ of 0.1 and a factor of 3 in viscosity because of the proximity to the critical point $\phi_c$ \citep{BJ09}.  The remarkable feature of Fig.~\ref{fig:gapsize_epsilon_spheres} is that $1-\epsilon$, characterizing the strength of shear thickening, is independent of the gap size all the way down to two particle diameters until jamming dominates the rheology.  Thus the transition from a strongly shear thickening suspension to one that is jammed is quite sudden at 2 particle diameters.

In this measurement series we reduced the gap down as far as  175 $\mu$m  ($1.97a$) where the normal stress applied on the sample reached 24 kPa, the maximum that can be applied by the rheometer.    In contrast, at 180 $\mu$m the normal stress remained below 20 Pa in the limit of zero  shear rate.   This suggests the yield stress of the suspension jumped dramatically between $1.97a$ and $2.02a$.  Despite the differences in procedure and packing fraction  from the measurements in Fig.~\ref{fig:gapsweep}, we again find the limiting gap size to be about $2a$.  We also note that for 180 and 190 $\mu$m gaps, or below about $2.1a$, the glass beads were audibly grinding underneath the plate when sheared.  When compared to the data in Fig.~\ref{fig:gapsweep}, this is the same region where the normal stress exceeds the shear stress by about an order-of-magnitude.  Both observations suggest the shear stress is dominated by friction in this regime.   Variations in this friction as particles grind past each other may explain why fluctuations occur in shear rate-controlled mode and why viscosity curves cannot be measured in stress-controlled mode.

\subsection{Rods}

\begin{figure}                                                
\centerline{\includegraphics[width=3.1in]{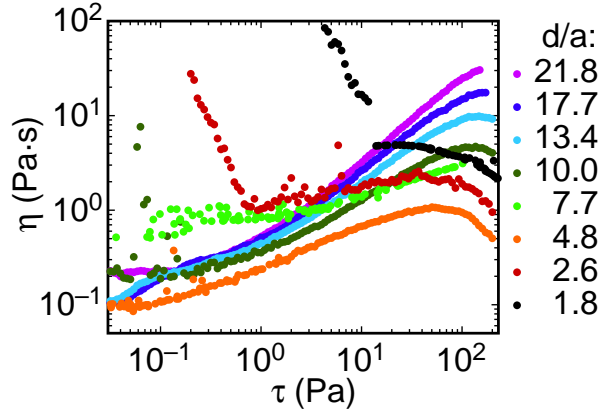}}
\caption{Viscosity curves for $\Gamma=9$ rods for different gap sizes at $\phi=0.33$. Gap sizes $d/a$ in units of particle widths are shown in the key.}
\label{fig:Gamma12}                                        
\end{figure}

\begin{figure}                                                
\centerline{\includegraphics[width=2.7in]{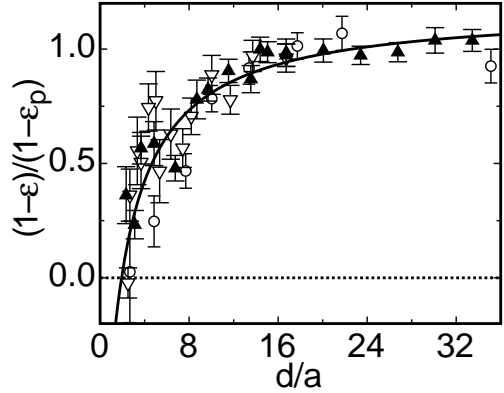}}
\caption{$1-\epsilon$ characterizing the logarithmic slope of the viscosity curve according to $\eta\propto \tau^{1-\epsilon}$ for rods with aspect ratio $\Gamma=9$.  The gap size $d/a$ is in units of particle width. Data are for several packing fractions and normalized by the value of $1-\epsilon_p$ corresponding to the logarithmic slope for bulk shear thickening for each packing fraction.    Open circles:  $\phi=0.33$, $1-\epsilon_p=0.84$.   Open triangles:  $\phi=0.32$, $1-\epsilon_p=0.58$.  Solid triangles:  $\phi=0.24$, $1-\epsilon_p = 0.28$. Dotted line:  $1-\epsilon = 0$ corresponds to a Newtonian scaling.  Solid line:  fit of Eq.~\ref{eqn:alphafit} to the data.  }
\label{fig:epsilonGamma12}                                        
\end{figure}

Now that we have this baseline result for spherical particles, we turn our attention to rod-shaped particles which have two length scales and whose structure can evolve as the gap size is reduced and rotation is suppressed.  In this case we performed all measurements in shear rate-controlled mode to avoid transient jamming.  Viscosity curves for rods of aspect ratio $\Gamma=9$ for different sample thicknesses are shown in Fig.~\ref{fig:Gamma12}.  A yield stress appears for small gaps, and at the smallest gaps corresponding to less than about two particle widths the suspension is jammed again in the measurement range of our rheometer, much like for spheres.  The logarithmic slopes $1-\epsilon$ are obtained from fits to the viscosity curves in the fixed stress range of 4-20 Pa which corresponds to the region of steepest slope, except for the smallest gap sizes where the yield stress encroaches on this region.  In this case, we measured the most positive slope to be sustained over a factor of 2 in stress.  These values of $1-\epsilon$ are plotted in Fig.~\ref{fig:epsilonGamma12}.    To check that the yield stress does not bias the slope, we refit $\eta \sim \tau^{1-\epsilon}$ with the yield stress subtracted off the data.   The fit values of $1-\epsilon$ did not shift beyond the given error bars.   Unlike for spheres, we see that the slope of the shear thickening curve decreases as the gap size $d$ decreases below a few particle lengths $\Gamma a$.  The slope gradally approaches zero, indicating that the shear thickening regime transitions into a Newtonian scaling regime in the limit of 2 particle widths.

Similar measurements of viscosity curves at small gap sizes were made for three different packing fractions of rods, all below $\phi_c$, and logarithmic slopes are shown in Fig.~\ref{fig:epsilonGamma12}.   In each case the slope $1-\epsilon$ is normalized by the limiting value $1-\epsilon_p$ obtained at large gap sizes.  The data for all three packing fractions tends to overlap when normalized in this way despite the fact that for bulk shear thickening the value of $1-\epsilon$ varies strongly with the packing fraction \citep{BJ09}, suggesting a universal function describes the transition from shear thickening to a Newtonian scaling regime under confinement.

\begin{figure}                                                
\centerline{\includegraphics[width=3.25in]{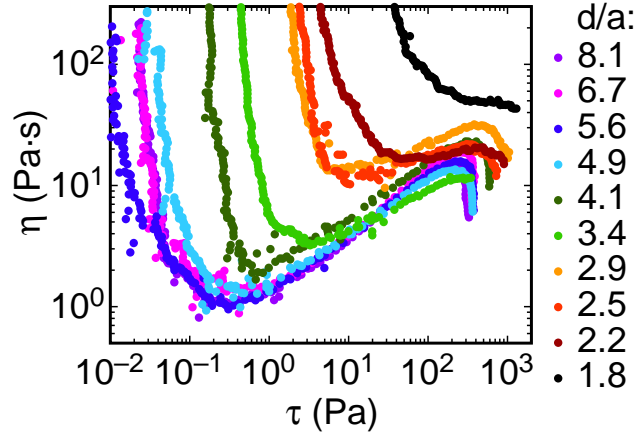}}
\caption{Viscosity curves for $\Gamma=6$ rods for different gap sizes at $\phi=0.31$. Gap sizes $d/a$ in units of particle widths are shown in the key.}
\label{fig:Gamma6}                                        
\end{figure}

\begin{figure}                                                
\centerline{\includegraphics[width=3.in]{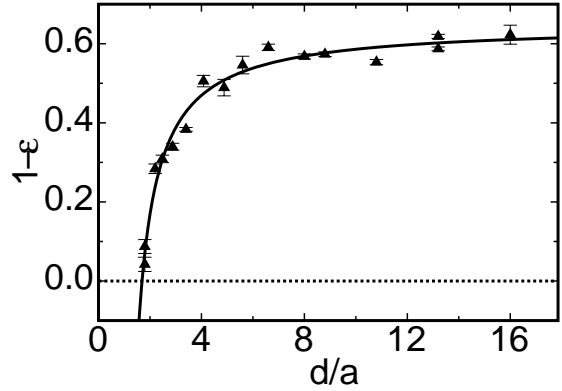}}
\caption{$1-\epsilon$ characterizing the logarithmic slope of the viscosity curve according to $\eta\propto \tau^{1-\epsilon}$ for rods with aspect ratio $\Gamma=6$.    The gap size $d/a$ is in units of particle widths.  Dotted line:  $\epsilon = 0$ corresponds to a Newtonian scaling.  Solid line:  fit of Eq.~\ref{eqn:alphafit} to the data. }
\label{fig:epsilonGamma6}                                        
\end{figure}

To determine how this transition depends on rod length, we did a similar set of measurements for $\Gamma=6$ rods, shown in Fig.~\ref{fig:Gamma6}. We fit the logarithmic slope in the stress range of 10-50 Pa again corresponding to the region of steepest slope.  If the yield stress encroached on this range, we measured the steepest slope as for the $\Gamma=9$ rods.  Since the yield stress becomes large at very small gap sizes and is accompanied by a shear thinning regime, in principle the measured slope could be a crossover regime rather than an indication of the shear thickening regime itself.  To check for such behavior, we additionally fit the stress/shear-rate curves in the initial shear thinning regime and shear thickening regime to the form 

\be
\tau = \tau_y+ a_1\dot\gamma^{\alpha}+a_2\dot\gamma^{1/\epsilon} \ .
\label{eqn:HBmod}
\ee

\noindent The first two terms characterize the shear thinning behavior by a Herschel-Bulkley form with $0<\alpha <1$ \citep{GZ04}.  The third term characterizes the shear thickening behavior with $0<\epsilon <1$ and is equivalent to $\eta\sim\tau^{1-\epsilon}$.  This linear sum characterizes the transition between shear thinning and shear thickening well for suspensions that exhibit both shear thickening and a yield stress \citep{BFOZMBDJ10}.  The values of $1-\epsilon$ obtained in this way are slightly higher by about $0.1$ in the cases where there is a large yield stress for $d\stackrel{<}{_\sim}2.5a$ than those by locally fitting $\eta \sim \tau^{1-\epsilon}$.   For these small gap sizes where there is a difference between the two fitting methods, we show the values obtained from fitting Eq.~\ref{eqn:HBmod} in Fig.~\ref{fig:epsilonGamma6}.  Using either fitting method, the value of $1-\epsilon$ decreases as the gap size decreases and goes to zero at close to 2 particle layers.  This similarity between the two fitting methods suggests that the leveling off of the viscosity curves is not primarily due to a crossover effect but instead is due to a change in the scaling of the underlying shear thickening regime.  It is apparent that the transition from shear thickening to Newtonian scaling occurs over a shorter range of gap size than for the longer rods, suggesting the range of the transition depends on the particle length.

\subsection{Characteristic lengthscales}

We can identify two length scales in the transition from a shear thickening to a Newtonian scaling shown in Figs.~\ref{fig:epsilonGamma12} and \ref{fig:epsilonGamma6}.  One is the gap size at which $1-\epsilon=0$, corresponding to Newtonian scaling.  We will represent this scale by $d_0$.   The other is the characteristic width of the transition which we will represent by $\Lambda$.  We desire a fitting function from which we can obtain these length scales as well as capture the plateau at the value $1-\epsilon_p$ in the limit of large gap size.  The simplest fitting function that has these features is 

\be
1-\epsilon(d) = \frac{1-\epsilon_p}{1+(\Lambda-d_0)/(d-d_0)} 
\label{eqn:alphafit}
\ee

\begin{figure}                                                
\centerline{\includegraphics[width=3.in]{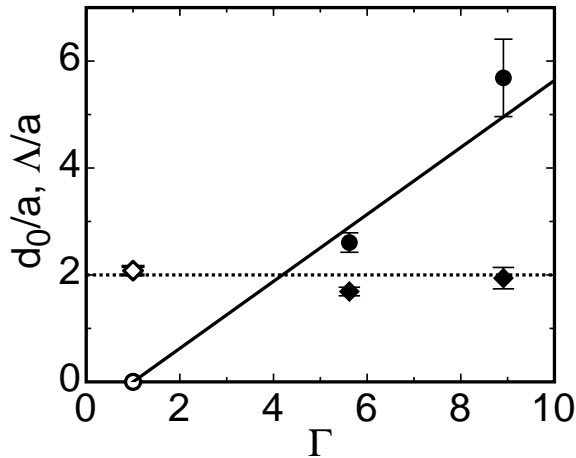}}
\caption{The rheological length scales $\Lambda$  (circles) corresponding to the gap size range over which shear thickening evolves into a Newtonian scaling regime and $d_0$ (diamonds) corresponding to the gap size where the Newtonian limit is reached.  These are normalized by the particle width $a$ and are shown for spheres (open symbols) and rods (solid symbols) as a function of  aspect ratio $\Gamma$.  Dotted line $d_0/a = 2$.  Solid line: fit of $\Lambda/a \propto (\Gamma-1)$.}
\label{fig:crossoverGamma}                                        
\end{figure}

\noindent To obtain the characteristic rheological length scales from the slope measurements, we  fit Eqn.~\ref{eqn:alphafit} to the data for rods of each $\Gamma$.  For this fitting function, $\Lambda$ is precisely the gap size where the slope is half the plateau value, and a linear extrapolation from $d=d_0$ at a constant slope crosses $1-\epsilon_p$ at $d=\Lambda$.  For the $\Gamma=9$ data, since we have three different packing fractions we normalized each by the plateau value of $1-\epsilon_p$ as shown in Fig.~\ref{fig:epsilonGamma12} so we can fit all three data sets to the same function.  For $\Gamma=9$ we obtain $d_0 = 58\pm6$ $\mu$m and $\Lambda = 170\pm20$ $\mu$m.  For $\Gamma=6$ we obtain $d_0 = 42\pm 1$ $\mu$m and $\Lambda=64\pm3$ $\mu$m.  These uncertainties are the statistical uncertainties from the fit corresponding to one standard deviation.  These fit values normalized by particle widths are shown in Fig.~\ref{fig:crossoverGamma} as a function of $\Gamma$.  Also shown is the value of $d_0$ for the transition to jamming for spheres which can be considered to have an aspect ratio $\Gamma=1$.  Since spheres showed a sudden transition from bulk shear thickening to jamming (i.e. there was no gradual transition to a Newtonian scaling), this corresponds to $\Lambda=0$.  Based on how close we measured to $d_0$ and our uncertainties, we can put an upper bound on $\Lambda$ of 0.01 for spheres.  All three value of $d_0$ are close to $2a$ and consistent with the range $2a-\delta a < d_0 < 2a$, indicating that $d_0$ is set by the particle width.  The value of $\Lambda$ is on the same order as the particle length for rods and seems to scale with the particle length.  Since $\Lambda =0$ for $\Gamma=1$ spheres, this suggests the crossover should perhaps scale $\Gamma-1$ instead of $\Gamma$.  A fit of the scaling $\Lambda/a \propto \Gamma-1$ is shown in Fig.~\ref{fig:crossoverGamma}.  This linear function is seen to fit the data within about a standard deviation, confirming that the width of the transition from a shear thickening to Newtonian scaling regime is consistent with a dependence on $\Gamma-1$.    Note that since $\Lambda$ is a crossover scale, its magnitude is somewhat arbitrary in nature and so it is not meaningful to quantify the proportionality beyond the order-of-magnitude.  Since $\Lambda$ is around the size of particle length it seems  the  transition from a shear thickening to Newtonian scaling regime coincides with the suppression of the rotational degree of freedom of particles by confinement.

\section{Discussion and Conclusions}


We have measured three length scales (the wavelength of oscillations $\lambda$, the Newtonian limit $d_0$, and the crossover scale $\Lambda$ from a shear thickening to Newtonian scaling regime) that show up in the rheology of confined suspensions and connected them directly to the dimensions of the particles in the suspensions.  Spheres or rods compressed down to two particle widths ($=d_0$) were found to jam in every experiment.   Spheres and rods also showed confinement-induced oscillations at small gap sizes with a wavelength $\lambda$ commensurate with the particle width as seen in Fig.~\ref{fig:gapsweep}.   The major differences in rheology for different particle shapes occurred at gap sizes between two particle widths and a few particle lengths.  The kink seen in Fig.~\ref{fig:gapsweep}b as the suspension of rods was reduced below one particle length indicates the onset of extra flow resistance from the interaction between particles and boundaries.   This behavior is also seen in the viscosity curves of Fig.~\ref{fig:Gamma12} and Fig.~\ref{fig:Gamma6} as the viscosity at low stress increases due to the appearance of a yield stress at smaller gap sizes.  As the the gap is reduced below about a rod length, the number of degrees of freedom for the particles is reduced as they may no longer fully rotate out of the plane made by the shear direction and gradient.    Long rods also showed a gradual transition of width $\Lambda$ from a shear thickening regime to a Newtonian scaling regime in the limit of $d_0$.  The width of the transition could be characterized by a dependence on aspect ratio $\Gamma-1$.  This supports an interpretation based on the freedom of particle rotation since $\Gamma-1=0$ corresponds to a shape which has no restrictions on particle rotation and $\Gamma-1$ can be considered a measure of how much free space is required for a particle to rotate and thus how large a gap is required to reach the bulk shear thickening limit.   The fact that the transition from a shear thickening to Newtonian scaling regime coincides with the suppresion of the rotational degree of freedom of particles suggests that this degree of freedom may be necessary for Discontinuous Shear Thickening.

\begin{figure}                                                
\centerline{\includegraphics[width=3.in]{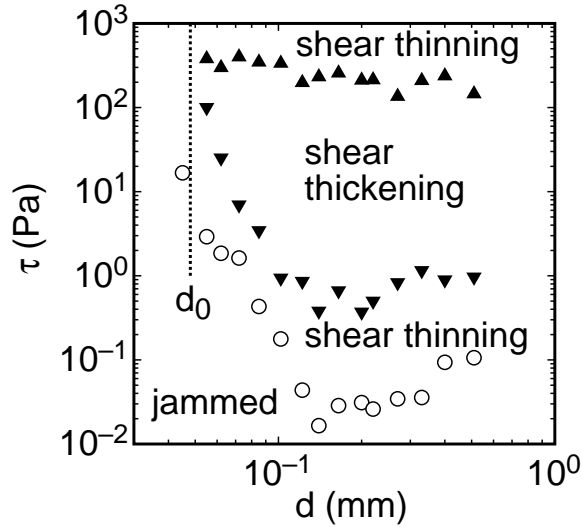}}
\caption{A non-equilibrium phase diagram showing shear thickening, shear thinning, and jammed regions as a function of gap size for $\Gamma=6$ rodsd at $\phi=0.31$.  Open circles:  yield stress.  Down-pointing triangles: onset of shear thickening.  Up-pointing triangles:  maximum stress of the shear thickening regime.   Dotted line: $d_0$ which corresponds to the point where the shear thickening slope $1-\epsilon$ goes to zero. }
\label{fig:phasegapsize}                                        
\end{figure}

There is an interesting contrast between the shear thickening phase boundaries due to confinement and due to jamming at $\phi_c$.  For each shape, as the gap size was reduced and the yield stress increased, it encroached on the shear thickening regime which usually remains in a relatively fixed stress range without the influence of a yield stress.  When the confinement leads to a large enough yield stress below a gap about two particle widths across, the shear thickening regime no longer exists.  This is consistent with the picture that a yield stress from any source will hide shear thickening behavior and eliminate it if the total shear thinning stress exceeds the upper stress range of the shear thickening regime \citep{BFOZMBDJ10}.  This was shown to be the case regardless of the source of the yield stress: it could be due to either chemical or induced attractions, as well as from confinement when the packing fraction exceeds the jamming point.  Here we have shown that another source of confinement, namely bringing the walls closer together, leads to the same effect.  The non-equilibrium phase diagram showing the shrinking of the shear thickening stress range as the yield stress increases at smaller gap sizes is shown in Fig.~\ref{fig:phasegapsize}.  It shows that the yield stress and accompanying shear thinning regime shrink the shear thickening regime to zero at about $d_0$.   An interesting contrast with jamming in the bulk can be made by noting that the introduction of a yield stress by confinement shrinks and eliminates the shear thickening regime just as ordering of the rods transforms the shear thickening regime into a Newtonian scaling regime.  While Discontinuous Shear Thickening is eliminated by confinement either by increasing the packing fraction or decreasing the gap size, the shear thickening behavior approaches two different limits.  As $\phi\rightarrow \phi_c$ from below, $\epsilon\rightarrow 0$, corresponding to a discontinuous jump in stress with shear rate (the strongest possible shear thickening), regardless of shape \citep{BJ09,EW05}.  On the other hand as $d\rightarrow d_0$ from above for rods, $\epsilon\rightarrow 1$, corresponding to a Newtonian scaling regime.  This difference may be expressed in terms of the order of the system; while $\phi\rightarrow \phi_c$ from below corresponds to the epitome of disorder \citep{OSLN03}, $d\rightarrow d_0$ from above corresponds to a forced planar ordering for rods.  The fact that bulk Discontinuous Shear Thickening is qualitatively similar for both spheres and rods [see  Figs.~\ref{fig:gapsize_spheres},\ref{fig:Gamma12},\ref{fig:Gamma6}, and \citet{EW05}] but the spheres do not have a transition to a Newtonian scaling regime underscores the fact that this difference is the result of forced ordering of rods at small gaps.

We showed that confined suspensions over a wide range of packing fractions will jam when compressed down about 2 particle layers (Fig.~\ref{fig:jamphase_gap}).  This transition from fluid-like behavior to jamming is similar to the case for pure fluids which also generate a frictional resistance to shear when they are compressed down to two molecules thick \citep{VG88}.  Confinement-induced oscillations have also been observed in channel flow where the channel width becomes comparable to the particle spacing \citep{ASJBKV00}.  While we have observed these confinement-induced oscillations for gap sizes up to 9 particle diameters for spheres, Discontinuous Shear Thickening remains largely unaffected in this range.  Therefore, these commensurability effects of confinement gave no indication that they were related to shear thickening and we suspect they are general features of the granular limit of fluids.


It has been suggested that shear thickening is a form of transient jamming \citep{FMB97, CWBC98}.  However, the transient jamming observed in Fig.~\ref{fig:smallgap_control} should be considered distinct from Discontinuous Shear Thickening.  The latter is a reproducible function of shear rate or stress and has viscosity curves independent of the control mode.  From Fig.~\ref{fig:smallgap_control} it is apparent that the underlying viscosity curve without these jumps is still shear thickening with the same slope as at larger gap sizes.  Thus the shear thickening observed at larger gaps cannot be the result of a smoothing of these transient jamming events over a larger sample.  Despite the fact that there is transient jamming for small gap sizes, the underlying shear thickening behavior was not affected at all for spheres until the permanent jamming transition eliminated it.  Thus transient jamming of the type we observed was an additional effect of confinement and not the cause of Discontinuous Shear Thickening.  We note that \citet{CZ95} found a qualitatively similar effect, in which a weak shear thickening effect associated with large fluctuations of the apparent viscosity emerged as the gap size was reduced below about 50 particle layers.  


We found that rods were forced to align parallel to the plates at small gaps at which shear thickening was eliminated (see Figs.~\ref{fig:epsilonGamma12} and \ref{fig:epsilonGamma6}).  This dependence on particle arrangement may be relevant for certain mechanisms of stress transmission to the rheometer plates.  For example, purely viscous lubrication should occur equally well for rods aligned with the plates, but since we find shear thickening disappears those sorts of couplings are not likely responsible for Discontinuous Shear Thickening.  On the other hand, if the coupling involves some compressional component, long rods forced to align with the shear cannot easily pile up to transfer shear and compressive stress to the plates.  We also observe an increasing upward normal force on the rheometer tool as the shear rate increases in the shear thickening regime, and that normal stress is typically on the same order as the shear stress.  These observations suggest that the mechanism stress transfer to the rheometer plates for Discontinuous Shear Thickening is compressive and not viscous.

One proposed mechanism in which compressional stress contributes to shear thickening is hydroclustering \citep{BB88, FMB97, SWB03, MB04b} in which the compressive and viscous lubrication stresses cause particles to cluster along the compressive axis at a 45$^{\circ}$ diagonal between the shear and shear gradient directions.  Our observation that shear thickening is eliminated for rods may be expected for such a mechanism because in such a layered arrangement they cannot easily align diagonally to transmit stress along the compressive axis.  Another proposed mechanism for shear thickening is dilation \citep{MW58, Ba89, LDHH05, FHBOB08}.  Densely packed hard spheres must dilate under shear because their shape does not allow them to naturally arrange into layers that can shear over each other without interference.  We note that we also observe an apparent increase in the roughness of the suspension  surface as the shear rate increases in the shear thickening regime.  Dilation requires at minimum two particle layers -- one to ride over the other -- and this is exactly what was found in Fig.~\ref{fig:gapsize_epsilon_spheres} as the minimum requirement for Discontinuous Shear Thickening.   Rods can also dilate when sheared in disorganized structures because some particles have rotated out of alignment to block the motion of other particles.  In contrast, when rods are forced to align in small gaps they can arrange into organized layers that do not have any overlap.   These layers do not need to dilate when sheared and this is precisely the condition under which we find Discontinuous Shear Thickening to weaken and eventually disappear. 

For either mechanism, it is still unclear how to explain the steep stress/shear-rate relations characteristic of Discontinuous Shear Thickening.  Our findings on how the elimination of shear thickening under confinement depends on particle geometry may provide additional constraints to resolve this problem.  Additionally, our finding that shear thickening behavior scales down to as few as two particle layers for spheres may  also offer advantages for simulations in which it is difficult to model complicated particle interactions for large numbers of particles.

\section{Acknowledgements}

This work was supported by DARPA through US Army Research Office grant W911NF-08-1-0209.  We acknowledge the NSF MRSEC program under DMR-0820054 for the use of shared equipment.

\section{Appendix: The shear profile and slip}
\label{sec:slip}

A variation of apparent viscosity with gap size in rheological measurements has often been attributed to wall slip.  Slip can result in a higher measured shear rate than the average shear rate $\dot\gamma_f$ in the fluid.  It is usually characterized by a slip length $l_{s}$ of the plate displacement in excess of the fluid displacement next to the plate per unit strain of the fluid.  This gives a measured shear rate of $\dot\gamma = \dot\gamma_{f}(1+2l_{s}/d)$  \citep{YP88, MW01a, EW05}.  This would tend to give a lower measured viscosity for small gap sizes with a more dramatic effect for $d < l_s$.  Slip has been measured to have an increasing effect at higher applied stresses and packing fractions \citep{CBP05,EW05}.   This leads to less steep shear thickening curves at higher stresses, and can even cause curves at higher packing fractions to appear less steep than curves at lower packing fractions \citep{CBP05}, contrary to the usual trend for Discontinuous Shear Thickening \citep{MW01a, EW05, BJ09}.   Some features of our measurements are qualitatively different from slip, for example the slopes of the viscosity curves do not decrease at higher stresses within the shear thickening regime, and the weakening of shear thickening at small gap sizes for rods does not go away at lower packing fractions as expected for slip.   However, the general trend of lower viscosity at smaller gap sizes is qualitatively similar to our observations for rods, and even the functional dependence on $d$ is similar to what we find in Figs.~\ref{fig:epsilonGamma12} and \ref{fig:epsilonGamma6} (although the effect of slip is expected to be on the viscosity magnitude instead of the slope).  Since the traditional way to determine slip is based on measuring the dependence of the apparent viscosity on gap size \citep{YP88}, a more direct measurement of slip is needed to distinguish between slip and granular confinement effects at small gaps so we can confirm that the measured gap size dependence was not a result of slip.

\begin{figure}                                                
\centerline{\includegraphics[width=3.25in]{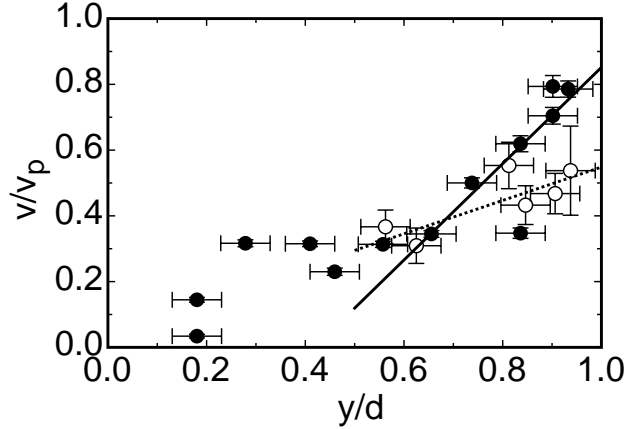}}
\caption{ Reconstructed shear profile for individual $\Gamma=9$ rods tracked using video microscopy.  Velocity $v$ normalized by plate velocity $v_p$ as a function of height $y$ normalized by plate height $d$. Solid circles:  $d=0.52$ mm.  Open circles:  $d=0.096$ mm.  Solid line: fit of $v/v_p = (1+2l_s/d)^{-1} -a_1 (1-y/d)$ to solid circles for $y>d/2$ to estimate the slip length $l_s$. Dotted line:  fit of the same function to open circles.  The amount of slip found is too small to account for the transition from a shear thickening to Newtonian scaling regime for rods.}
\label{fig:shearprofile}                                        
\end{figure}

To measure slip directly we measured shear profiles optically in some experiments.   We used a video camera with a microscope lens that can resolve  20 $\mu$m.  The camera was placed next to the rheometer and focused on a 4 mm across by 1 mm high area in the plane of the shear direction and shear gradient at the outer edge of the sample.  While there is some distortion from looking through the curved liquid-air interface, these videos can observe individual particle motions and can be used to estimate the shear profile.  From these videos we observed that the shear flow is fully developed with an approximately linear shear profile at stresses above the shear thickening regime.   At lower stresses in the shear thickening regime the velocity gradient is not uniform in gap size but higher near the moving plate.  This is the case for both spheres and rods in qualitative agreement with MRI measurements of cornstarch \citep{FHBOB08}.  It appears as if there is a shear band near the top plate and that there is only minimal net motion further down in the layer.  The width of this effective shear band decreases as the stress decreases, and below about the onset of shear thickening, there is almost no particle motion at all when the particles are denser than the liquid because they settle on the bottom plate.  This settling occurs in the steady state limit regardless of shear history because the shear stress is not enough to keep the particles suspended.  This suggest that some of the nonlinearity at low stresses can be due to gravity, although a significant nonlinearity is still observed in density matched cornstarch in a Couette geometry \citep{FHBOB08}.  We also observed for rods that while most particles tend to be aligned with the shear flow, many rotate out of the horizontal plane where they have the space to at larger gap sizes.  Rotations all the way around by 180 degrees are found even for $d < \Gamma a$ where they cannot be fully in a vertical plane but must be slanted somewhat out of the plane of view.  From these videos we manually tracked the horizontal distance traveled by individual rods over time intervals of up to 160 s (sometimes less time if the particle could not be reliably tracked).  The mean velocity $v$ of each particle was plotted as a function of the mean height $y$ of that particle over the tracking period in Fig.~\ref{fig:shearprofile}.  The velocities and heights were normalized by the plate velocity $v_p$ and plate height $d$ which were calibrated by the videos.  To estimate the slip length, we note that $v(y=d)/v_p = \dot\gamma_f/\dot\gamma = (1+2l_s/d)^{-1}$.  We then fit a first order expansion $v/v_p = (1+2l_s/d)^{-1} -a_1 (1-y/d)$ to this data for $y >d/2$ to obtain $l_s$.   We do not mean for this to be an exhaustive study of the shear profile or imply  the shear profile is linear by this fit.  For $\Gamma=9$ rods at $\phi=0.29$ (below $\phi_c$) and $\dot\gamma=0.1$ s$^{-1}$ (in the shear thinning regime at stresses above shear thickening) we did this analysis for two different gap sizes.  For $d=520$ $\mu$m, this gives a slip length of $l_s  = 50\pm 30$ $\mu$m.  For $d=96$ $\mu$m, this gives a slip length of $l_s = 39\pm 4$ $\mu$m.  The data for the two gap sizes give a consistent value for the slip length as expected.   The shear rate measured at the smaller gap size was an order of magnitude smaller than for a bulk suspension at the same stress.  However, the value of the slip length measured could explain at most a measured shear rate ratio of 1.6 between the two gap sizes.  This value is the worst case scenario if the functional form of slip is the same as the geometric effects we observed; as explained in the previous paragraph there are also some qualitative differences.   In fact, the maximum possible shear rate ratio, which occurs in the limit of large slip lengths is equal to the ratio of the gap sizes.  So even for very large slip lengths the ratio of measured shear rates due to slip could never be greater than 5.4.  Alternatively, the notable effect of the gap size on Discontinuous Shear Thickening is the change in the slope of the viscosity curve.   From that point of view, one should ask if the increase in shear rate with stress in the shear thickening regime can be the result of slip.  By comparison, in the bulk behavior at high packing fractions the shear rate is approximately constant in the shear thickening regime. For $\Gamma=9$ and $d=96$ $\mu$m the shear rate increases by about a factor of 10 (see Fig.~\ref{fig:Gamma12}) again this is too large an effect to be explained by slip.  

Therefore, considering the smallness of the measured values of slip, and the lack of a self-consistent slip model that could explain the variation in the slope of the viscosity curves with gap size shown in Figs.~\ref{fig:epsilonGamma12} and \ref{fig:epsilonGamma6}, we conclude that these results cannot primarily be due to slip. Additionally, since the gap-size dependence at few particle layers is not consistent with the traditional slip correction models based on the hydrodynamic limit \citep{YP88}, then we cannot use this gap size dependence to correct for slip effects.

\end{document}